\documentclass[%
reprint,
superscriptaddress,
showpacs,preprintnumbers,
nofootinbib,
amsmath,amssymb,
aps,
prd,
floatfix,
]{revtex4-1}

\usepackage{graphicx}
\usepackage{dcolumn}
\usepackage{bm}
\usepackage{hyperref}
\usepackage[mathlines]{lineno}
\usepackage{float}
\usepackage{color}

\begin{document}


\title{
Pitfalls of a power-law parameterization of the primordial power spectrum for Primordial Black Hole formation
}

\author{Anne M. Green}
\email{anne.green@nottingham.ac.uk}
\affiliation{School of Physics and Astronomy, University of Nottingham, University Park, Nottingham, NG7 2RD, United Kingdom}
\date{\today}
\begin{abstract}
Primordial Black Holes (PBHs) can form in the radiation dominated early Universe from the collapse of large density perturbations produced by inflation. A power-law parameterisation of the primordial power spectrum is often used to extrapolate from cosmological scales, where the amplitude of the perturbations is well-measured by Cosmic Microwave Background and Large Scale Structure observations, down to the small scales on which PBHs may form. 
We show that this typically leads to large errors in the amplitude of the fluctuations on small scales, and hence extremely inaccurate calculations of the abundance of PBHs formed.
\end{abstract}

\maketitle

\section{Introduction}

    The recent discovery of gravitational waves from $\sim 10 \,  M_{\odot}$ binary BH mergers has led to a resurgence of interest in Primordial (i.e.~formed in the early Universe) Black Holes (PBHs)~\cite{Bird:2016dcv,Carr:2016drx}. The abundance of planetary to multi-Solar mass PBHs is tightly constrained by microlensing~\cite{Allsman:2000kg,Tisserand:2006zx,Mediavilla:2017bok,Zumalacarregui:2017qqd}, dynamical~\cite{Monroy-Rodriguez:2014ula,Brandt:2016aco,Koushiappas:2017chw} and accretion~\cite{Ali-Haimoud:2016mbv,Gaggero:2016dpq,Inoue:2017csr} constraints. However there is a mass window at $(10^{-14}-10^{-10}) \, M_{\odot}$ where PBHs could make up all of the dark matter\footnote{Constraints have been published in this region from microlensing~\cite{Niikura:2017zjd}  and the destruction of neutron stars~\cite{Capela:2013yf}. However it has been pointed out that the standard microlensing analysis is not valid below $10^{-10} \, M_{\odot}$, since the wavelength of the light is larger than the PBH Schwarzschild radius and diffraction reduces the maximum magnification~\cite{Inomata:2017vxo}. The neutron star constraints meanwhile assume an unrealistically high dark matter density in globular clusters~\cite{Kawasaki:2016pql}.}~\cite{Carr:2016drx}.  PBHs can form from the collapse of large density perturbations~\cite{Carr:1974nx,Carr:1975qj} and avoiding PBH over-production constrains the primordial power spectrum on small scales, and hence models of inflation~\cite{Carr:1993aq}.
For a recent review of PBH formation and abundance constraints see Ref.~\cite{Sasaki:2018dmp}.

A non-negligible number of PBHs will only be formed from the collapse of density perturbations if their amplitude on small scales is several orders of magnitude larger than the measured amplitude on cosmological scales (e.g. Refs.~\cite{Carr:1993aq,Josan:2009qn,Bringmann:2011ut}). There are several ways in which this can be achieved. The amplitude of the fluctuations can grow smoothly with decreasing (physical) scale. This can occur, for instance, in the running mass inflation model~\cite{Stewart:1996ey,Stewart:1997wg,Leach:2000ea} and also a sub-set of the models produced by the flow formalism~\cite{Peiris:2008be,Josan:2010cj} (see Sec.~\ref{subsec:flow}). Another possibility is a spike or broad peak in the power spectrum, which can be produced by a feature in the inflaton potential~\cite{Motohashi:2017kbs,Ballesteros:2017fsr} or in multi-field models (see Ref.~\cite{Sasaki:2018dmp}).

It is common (c.f.~Ref.~\cite{Kosowsky:1995aa,Ade:2015lrj}) to parameterise the power spectrum of the primordial curvature fluctuations, ${\cal P}_{\cal R}(k)$, as a power-law:
\begin{equation}
\label{plps}
{\cal P}_{\cal R}(k) = A_{\rm s} \left( \frac{k}{k_{0}} \right)^{n_{\rm s}(k) -1} \,, 
\end{equation}
where $A_{\rm s}$ is the amplitude at the pivot wave-number, $k_{0}$, and 
\begin{equation}
\label{ns}
n_{\rm s}(k) = n_{\rm s} |_{k_{0}} + \alpha_{\rm s} \ln{\left( \frac{k}{k_{0}} \right)} 
  +  \beta_{\rm s}  \ln^2 {\left( \frac{k}{k_{0}} \right)} + ... \,,
\end{equation}
with
\begin{eqnarray}
\alpha_{\rm s} & = & \left. \frac{1}{2} \left( \frac{{\rm d} n_{\rm s}}{{\rm d} \ln{k}} \right) \right \rvert_{k_{0}}  \,,\\
\beta_{\rm s} &=& \left.  \frac{1}{6} \left( \frac{{\rm d}^2 n_{\rm s}}{{\rm d} (\ln{k})^2} \right) \right \rvert_{k_{0}} \,,
\end{eqnarray}
i.e.~the spectral index, $n_{\rm s}(k)$, is Taylor expanded around $k_{0}$. For Planck the pivot wave-number, $k_{0}$, was taken to be $0.05 \, {\rm Mpc}^{-1}$~\cite{Ade:2013zuv,Ade:2015lrj} as this is roughly in the middle of the logarithmic range of scales probed~\cite{Ade:2013zuv}~\footnote{From the perspective of accurately constraining the primordial power spectrum, the optimal choice of pivot scale is that corresponding to the multipole where the observational errors on the temperature power spectrum are smallest~\cite{Leach:2002ar}.}.

It is possible for inflation to produce a pure power-law power spectrum, with $\alpha_{\rm s}$, $\beta_{\rm s}$ and all higher order terms in the expansion of  $n_{\rm s}(k)$ identically equal to zero. However this only occurs for very specific forms of the inflaton potential~\cite{Mollerach:1993sy,Gilbert:1995wg,Vallinotto:2003vf}. If the primordial perturbations are produced by slow-roll inflation then generically $(n_{\rm s}-1)  \sim {\cal O} (\epsilon) \,,  \alpha_{\rm s} \sim {\cal O} (\epsilon ^2)$, $\beta_{\rm s} \sim {\cal O} (\epsilon^3)$ and so on, where $\epsilon < 1$ (e.g.~Ref.~\cite{Lidsey:1995np}). In this case the Taylor expansion of the spectral index is valid for cosmological observations, which probe a limited range of scales, $k \sim (10^{-4}-1) \, {\rm Mpc}^{-1}$. However this expansion has also been used when studying PBH formation (e.g.~Refs.~\cite{Alabidi:2009bk,Carr:2017edp,Kohri:2018qtx}), which occurs over a large range of much smaller length scales.
 The lightest PBHs which do not evaporate by the present day have $M_{\rm PBH} \approx 5 \times 10^{14} \, {\rm g}$~\cite{MacGibbon:1991vc}, which corresponds to a scale $k \sim  2 \times 10^{16} \, {\rm Mpc}^{-1}$, i.e.~$\ln(k/k_{0}) \sim 41$, while $M_{\rm PBH} \approx 10 \, M_{\odot}$ PBHs correspond to $k \sim 10^{6} \, {\rm Mpc}^{-1}$,  i.e.~$\ln(k/k_{0}) \sim 17$ (see Sec.~\ref{sec:background} for details). So for PBH formation $\epsilon \ln{(k/k_{0})}$ is not small, and therefore the power-law expansion in Eqs.~(\ref{plps}-\ref{ns}) is not expected to converge, even for slow-roll inflation models. If PBHs are formed from a spike or peak in the power spectrum on small scales, then a power-law extrapolation of the power spectrum from cosmological scales is clearly not appropriate.

In this paper we investigate the errors induced by using a power-law parameterisation of the power spectrum when studying PBH formation.
In Sec.~\ref{sec:background} we review the formation of PBHs from large density perturbations during radiation domination. In Sec.~\ref{sec:results} we study two cases, a Taylor expansion of the power spectrum truncated at different orders (Sec.~\ref{subsec:pl}) and inflation models generated using the flow formalism (Sec.~\ref{subsec:flow}). We conclude with discussion in Sec.~\ref{sec:discuss}.

\section{Background}
\label{sec:background}

In this section we outline how the abundance of PBHs depends on the amplitude of the primordial power spectrum. Since
our goal is not to carry out concrete calculations of the PBH abundance we do not consider, for instance, the dependence of the PBH mass on the amplitude of the density fluctuation due to critical collapse~\cite{Niemeyer:1997mt}, non-gaussianity of the primordial perturbations~\cite{Franciolini:2018vbk} or the uncertainties induced by the choice of window function~\cite{Ando:2018qdb}. For a detailed calculation see e.g. Refs.~\cite{Bringmann:2011ut,Sasaki:2018dmp}

During radiation domination a fluctuation on a physical scale $R$ will collapse to form a PBH, with mass $M_{\rm PBH}$ roughly equal to the horizon mass, $M_{\rm H}$, if the smoothed density contrast at horizon entry, $\delta(R)$, exceeds a threshold value $\delta_{\rm c}$ which is of order unity~\cite{Carr:1974nx,Harada:2013epa}. Assuming the initial perturbations have a Gaussian distribution then the initial PBH mass fraction, $\beta(M_{\rm PBH}) =  \rho_{\rm PBH}/\rho_{\rm tot}$, is given by~\cite{Carr:1975qj}\footnote{We follow the usual Press-Schecter procedure of multiplying the integral of the probability distribution by a factor of 2, so that all of the mass in the Universe is accounted for.}
\begin{eqnarray}
\label{beta}
\beta(M_{\rm PBH})  &\approx& \frac{2}{\sqrt{2\pi}\sigma(R)} 
\int_{\delta_{\rm c}}^{\infty} \exp{\left(- \frac{\delta^2(R)}
    {2 \sigma^2(R)}\right)} 
  \,{\rm d}\delta(R) \,,\nonumber \\
 &=&   {\rm erfc}\left(\frac{\delta_{\rm c}}{
   \sqrt{2}\sigma(R)}\right) \,,   \nonumber \\
   &\approx& \sqrt{ \frac{2}{\pi}} \frac{\sigma(R)}{\delta_{\rm c}} \exp {\left(-\frac{\delta^2(R)}
    {2 \sigma^2(R)}\right)}\,,
\end{eqnarray}
where $\sigma(R)$ is the mass variance evaluated when the scale of interest enters the horizon, ${\rm erfc}(x)$ is the complementary error function, and the last step uses its large $x$ approximation. 
The mass variance is given by~\cite{Sasaki:2018dmp}
\begin{equation}
\label{variance}
\sigma^2(R)= \frac{16}{81} \int_{0}^{\infty} W^2(kR) (kR)^4 {\cal P}_{\cal R}(k) \, {\rm d} \ln k \,,
\end{equation}
where ${\cal P}_{\cal R}(k)$ is the power spectrum of the primordial curvature perturbation on comoving slicing and $W(kR)$ is the Fourier transform of the window function used to smooth the density contrast. For a primordial power spectrum which varies slowly with scale $\sigma^2(R)$ is proportional to $\mathcal{P}_{\cal R}(1/R)$~\cite{Josan:2009qn,Bringmann:2011ut}. 

The observational constraints on the initial abundance of PBHs, $\beta(M_{\rm PBH})$,~\cite{Carr:2016drx,Sasaki:2018dmp} are scale dependent. However the PBH abundance depends exponentially on the mass variance, c.f.~Eq.~(\ref{beta}). Therefore the scale dependence of the resulting constraints on the amplitude of the primordial perturbations is relatively weak and the constraints can be roughly approximated as  $\mathcal{P}_{\cal R}(k) \lesssim 10^{-2}$~\cite{Josan:2009qn,Bringmann:2011ut,Ando:2018qdb}.

The horizon mass, $M_{\rm H}$, when a comoving scale $k$ reenters the horizon is given, using the Friedman equation and assuming radiation domination at early times (see Ref.~\cite{Motohashi:2017kbs,Ballesteros:2017fsr} for details), by 
\begin{equation}
M_{\rm H} =  5 \times 10^{14} \, {\rm g} \,  \left( \frac{g_{\star}}{106.75} \right)^{1/6} \left( \frac{ k }{2 \times 10^{16} \, {\rm Mpc}^{-1}} \right)^{-2} \,.
\end{equation}
The effective number of degrees of freedom, $g_{\star}$, has been assumed to be equal for entropy and energy density and normalized to its value at high temperatures in the Standard Model. Here we have normalised the horizon mass to the lightest PBHs which do not evaporate by the present day, $M_{\rm PBH} \approx 5 \times 10^{14} \, {\rm g}$~\cite{MacGibbon:1991vc}.

\section{Results}
\label{sec:results}

\subsection{Taylor expansion of $n_{\rm s}(k)$}
\label{subsec:pl}

We first examine how retaining a varying number of terms in the Taylor expansion of $n_{\rm s}(k)$, Eq.~(\ref{ns}), affects
the amplitude of the power spectrum on small scales. For concreteness we focus on a scale $k_{\rm ref}= 2 \times 10^{16} \, {\rm Mpc}^{-1}$ which corresponds, roughly, to the lightest PBHs which do not evaporate by the present day. For larger (smaller) $k$, corresponding to lighter (heavier) PBHs, the deviations in the power spectrum will be larger (smaller).

The constraints on the parameters of the power-law parameterisation of the power spectrum, Eq.~(\ref{plps}), including terms up to ${\rm d}^2 n_{\rm s}/{\rm d} (\ln{k})^2$,  from Planck 2015 using the TT, TE, EE+lowP data sets~\cite{Ade:2015lrj} are (with $1\sigma$ errors):
\begin{eqnarray}
\ln{(10^{10} A_{\rm s})} &=& 3.094 \pm 0.0034 \,,  \\
n_{\rm s} |_{k_{0}}  &=& 0.9586 \pm 0.0056 \,, \\
\left. \left( \frac{{\rm d} n_{\rm s}}{{\rm d} \ln{k}} \right) \right \rvert_{k_{0}}  &=& 0.009 \pm 0.010 \,,  \\
\left. \left( \frac{{\rm d}^2 n_{\rm s}}{{\rm d} (\ln{k})^2} \right) \right \rvert_{k_{0}} &=& 0.025 \pm 0.013 \,. 
\end{eqnarray}
For compactness we subsequently refer to $(n_{\rm s} |_{k_{0}} -1)$, $({\rm d} n_{\rm s}/{\rm d} \ln{k} ) |_{k_{0}}$
and $({\rm d}^2 n_{\rm s}/{\rm d} ( \ln{k})^2 ) |_{k_{0}}$ as the tilt, running and running of the running respectively.
We scan over the range of $2 \sigma$ allowed values of the tilt,  running and running of the running 
and calculate the power spectrum to 1st order (i.e.~using the tilt only and neglecting higher order terms), 2nd order (using the tilt and running)  and 3rd order (using the tilt, running and running of the running).
We find combinations of the parameters for which the 3rd order calculation gives $\mathcal{P}_{\cal R}(k_{\rm ref}) \approx 10^{-2}$ i.e. sufficiently large to form an interesting abundance of PBHs.

\begin{figure}[t]
\includegraphics[width=0.45\textwidth]{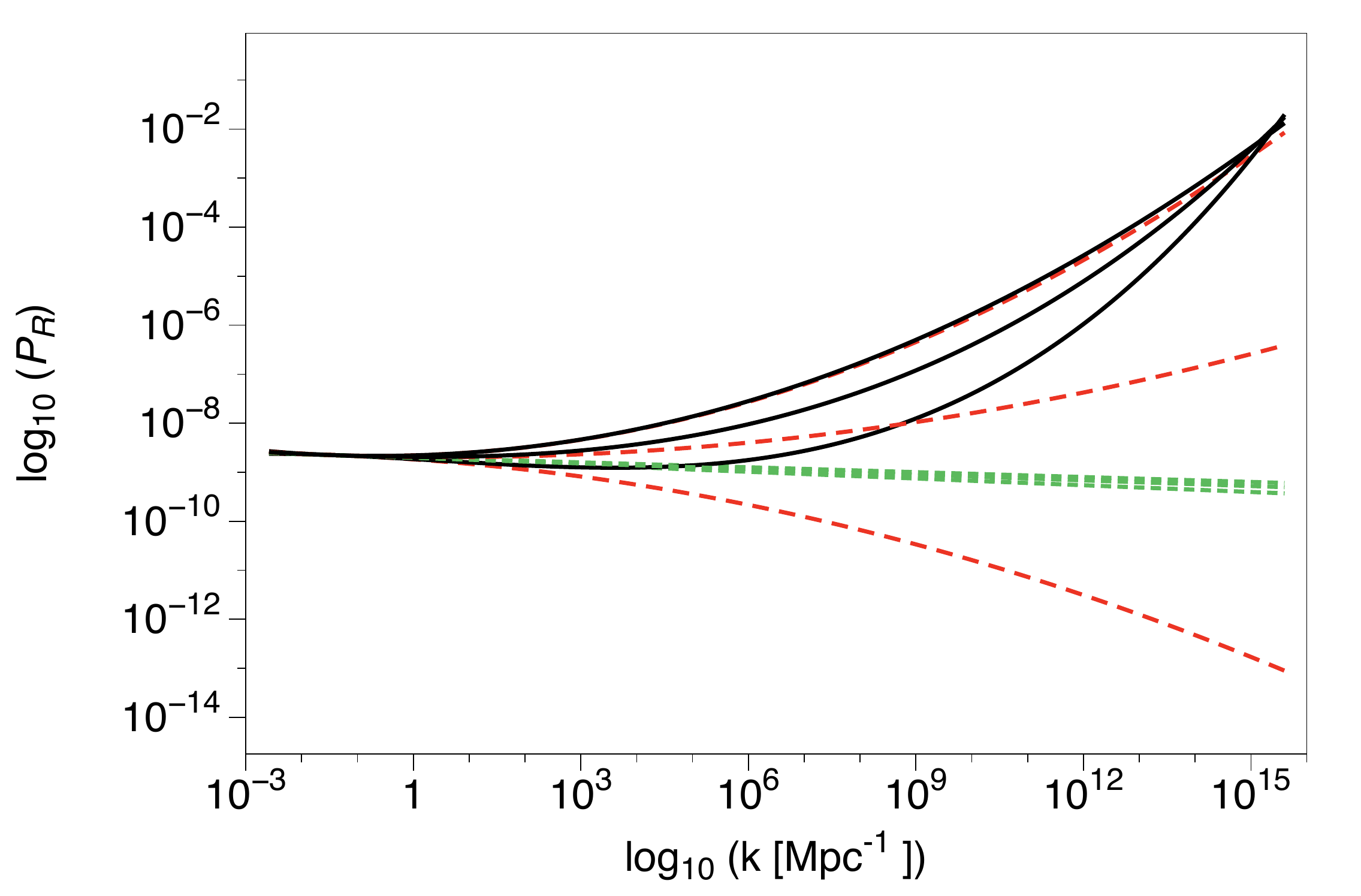}\\
\includegraphics[width=0.45\textwidth]{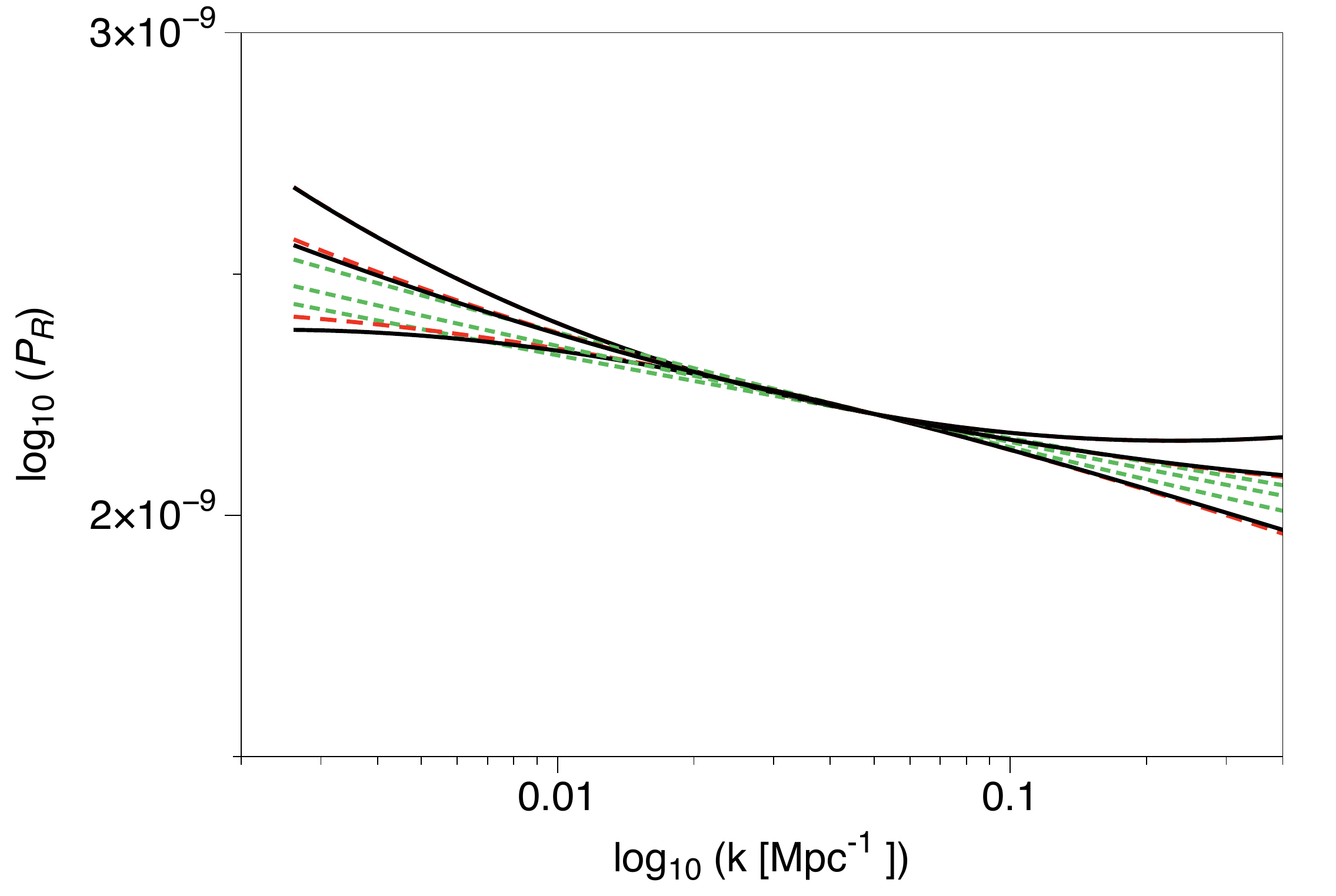}
\caption{\label{fig:PL} Three examples of the primordial power spectrum of the curvature perturbation, ${\cal P}_{\cal R}(k)$, calculated using a Taylor expansion of the scalar spectral index, $n_{\rm s}(k)$, retaining varying numbers of terms in the expansion. The black solid lines are for the 3rd order expansion including terms up to the running of the running. The red long-dashed and green short-dashed lines truncate the expansion at 2nd order (including the tilt and running) and 1st order (tilt only) respectively.  The tilt, running, and running of the running at the pivot scale, $k_{0}=0.05 \, {\rm Mpc}^{-1}$, have been chosen so that when the 3rd order expansion is used the perturbations on small scales are sufficiently large to form an interesting number of PBHs, specifically  ${\cal P}_{\cal R}(k_{\rm ref}= 2 \times 10^{16} \, {\rm Mpc}^{-1}) \approx 10^{-2}$, while also satisfying the constraints from Planck on large scales (see text for further details).  The upper panel shows the full range of scales considered while the lower panel is restricted to the scales constrained by Planck.}
\end{figure}

The 1st, 2nd and 3rd order calculations of the power spectrum are shown in Fig.~\ref{fig:PL} for three representative sets of parameters, chosen to span the range of variation. Since the tilt is small and negative the 1st order power spectra decrease weakly with increasing $k$ and are many orders of magnitude smaller on small scales than the 3rd order power spectra. The error on the tilt is small, so there is very little spread in the 1st order power spectra.
When the running of the running is allowed to vary in the fit to the CMB data, the allowed range of the running is quite large and encompasses (at $2 \sigma$) both positive and negative values. The variation in the 2nd order power spectra is therefore very large. $\mathcal{P}_{\cal R}(k_{\rm ref}) \approx 10^{-2}$ can be achieved with a positive running of the running and a negative running. In this case the 2nd order power spectrum decreases significantly with increasing $k$ while the 3rd order power spectrum increases significantly so that they differ on small scales by many orders of magnitude. $\mathcal{P}_{\cal R}(k_{\rm ref}) \approx 10^{-2}$ can also be achieved with a positive running and very small running of the running. In this case the 2nd and 3rd order calculations are in fairly good agreement, however this requires fine-tuning of the running of the running.

It is worth noting that when only the tilt and running are allowed to vary in the fit to the CMB data the errors on the running are significantly smaller, $({\rm d} n_{\rm s}/{\rm d} \ln{k} ) |_{k_{0}} = -0.0057 \pm 0.0071$ at $1\sigma$~\cite{Ade:2015lrj}. In this case the maximum allowed positive tilt and running (at $2 \sigma$) only produce $\mathcal{P}_{\cal R}(k_{\rm ref}) \sim 10^{-6}$, several orders of magnitude smaller than required to produce an interesting abundance of PBHs. 

We caution against over-interpreting these results, in particular they do not demonstrate that PBH formation is natural or generic. The power spectra can become large on small scales because of the large errors on the running and the running of the running when both are allowed to vary. These large errors reflect the fact that the running and running of the running are, to some extent, degenerate and the CMB observations do not probe a wide enough range of $k$ values to be sensitive to the running of the running. What this analysis does demonstrate is that using a power-law parameterisation of the power spectrum when studying PBH formation is inappropriate. The amplitude of the power spectrum varies by many orders of magnitude when the number of terms retained in the Taylor expansion of $n_{\rm s}(k)$ is changed.

\subsection{Flow formalism}
\label{subsec:flow}

The Taylor expansion of $n_{\rm s}(k)$ considered in Sec.~\ref{subsec:pl} makes no assumptions about the physics underlying the generation of the  primordial perturbations. We saw that the large uncertainties in the running and running of running, which arise because CMB observations probe a relatively limited range of scales, lead to significant differences in the power spectrum calculated when the expansion for $n_{\rm s}(k)$ is truncated at different orders.
However if the primordial perturbations are generated by slow-roll inflation then the running of the running is typically  smaller than the running, which in turn is smaller than the tilt~\cite{Lidsey:1995np}.
Therefore in this section we use the flow equations to examine how well a Taylor expansion of $n_{\rm s}(k)$ performs for a selection of inflation models where the power spectrum varies smoothly with scale and becomes large enough on small scales to produce an interesting density of PBHs.

The flow equations~\cite{Hoffman:2000ue,Kinney:2002qn} are expressed in terms of the
Hubble
slow-roll parameters~\cite{Salopek:1990jq},
\begin{eqnarray}
\epsilon_H &=&\frac{m_{{\rm{Pl}}}^2}{4\pi}\left(\frac{H^{\prime}(\phi)}{H(\phi)}\right)^2 \,, \\
^l\lambda_H  & = & \left( \frac{m_{{\rm{Pl}}}^2}{4\pi}\right)^l  \frac{(H^{\prime})^{l-1}}{H^l}
          \frac{{\rm d}^{(l+1)}H}{{\rm d}\phi^{(l+1)}} \hspace{5mm}  l\geq1 \,,
       \label{Hslowrollpara}
\end{eqnarray} 
where $m_{\rm{Pl}}$ is the Planck mass and ${}^\prime$ denotes
differentiation with respect to the scalar field, $\phi$, that drives inflation. The flow equations describe the variation of the slow-roll parameters 
in terms of the number of e-foldings from the end of inflation, $N=\ln{[a(t_{\rm{end}})/a(t)]}$:
\begin{eqnarray}
\frac{ {\rm d}\epsilon_H}{{\rm d}N} & = & \epsilon_H (\sigma_H+2\epsilon_H) \,,  \\         
\frac{{\rm d} \sigma_H}{{\rm d}N} & = & -5\epsilon_H \sigma_H -12 \epsilon_H^2+2(^2\lambda_H) \,, \\
\frac{{\rm d}(^l \lambda_H )}{{\rm d}N} & = & \left[ \frac{l-1}{2} \sigma_H + (l-2) \epsilon_H \right](^l\lambda_H) +^{l+1}\lambda_H \,,   \\
&& \hspace{50mm} l\ge 1               \nonumber
\label{flowequations}
\end{eqnarray}
where $\sigma_H=2(^1\lambda_H)-4\epsilon_H$.  Derivatives with respect to $N$ are related to derivatives
with respect to $\phi$ by
\begin{equation}
\frac{{\rm d} \,\,\,\,}{{\rm d} N} = \frac{m_{\rm Pl}}{2 \sqrt{\pi}} \sqrt{\epsilon_{\rm H}} \frac{{\rm d} \,\,\, \,}{{\rm d} N} \,,
\end{equation}
while
\begin{equation}
\label{kn}
\frac{{\rm d} \ln k}{{\rm d} N} = - (1- \epsilon_{\rm H}) \,.
\end{equation}

Kinney introduced the idea of using the flow equations to stochastically generate inflation models~\cite{Kinney:2002qn}.
He randomly chose `initial' values for the slow roll parameters and $N_{\rm{cos}}$, the number of e-foldings between
cosmological scales exiting the horizon and the end of inflation, in the ranges:
\begin{eqnarray}              \nonumber
N_{\rm cos} &=& [40, 70], \\ \nonumber
\epsilon_{H,0}&=&[0,0.8]  \,, \\   \nonumber
\sigma_{H,0}&=&[-0.5,0.5] \,, \\   \nonumber
^2\lambda_{H,0} &=&[-0.05,0.05]  \,,\\     \nonumber
^3\lambda_{H,0}&=&[-0.005,0.005]  \,, \\   \nonumber
&...& \\
^{M+1}\lambda_{H,0}&=&0 \,,
\label{hierarchy}
\end{eqnarray}
truncating the hierarchy at $M=5$, and used the flow equations to evolve the models forwards in time, until either inflation ended or a late-time fixed point was reached. The majority of the models generated by this procedure have a late time attractor with $\epsilon_{H} = 0$ where inflation is eternal and the amplitude of the perturbations becomes large. However this attractor is not always reached within $N_{\rm cos}$ e-foldings. We will follow Peiris and Easther~\cite{Peiris:2008be} and assume that an auxiliary mechanism (for instance a second scalar field) terminates inflation once the specified number of e-folding has occurred. Liddle~\cite{Liddle:2003py} pointed out that the flow equations effectively assume that the Hubble parameter is a polynomial function of the scalar field, $H(\phi)$, with the co-efficient of the $\phi^{m}$ term in $H(\phi)$ being determined by 
$(^{m-1}\lambda_{H,0})$ and $\epsilon_{H,0}$~\cite{Ramirez:2005cy}. Therefore only $\phi(N)$ (and in our case also $k(N)$) 
needs to be evolved numerically.

We use the same range of initial values for the slow-roll parameters as Ref.~\cite{Kinney:2002qn}, given in Eq.~(\ref{hierarchy}) above. For concreteness we fix $N_{\rm cos}=42$ so that the scale that exits the horizon at the end of inflation is, roughly (the relationship between $k$ and $N$, Eq.~(\ref{kn}), depends on $\epsilon_{\rm H}$), $k_{\rm ref} \approx 2 \times 10^{16} \, {\rm Mpc}^{-1}$.
We assume that $N_{\rm cos}=42$ corresponds to the Planck pivot scale $k_{0} = 0.05 \, {\rm Mpc}^{-1}$. We evolve each model 
forward in time (${\rm d}N<0$) from $N=42$ to $N=0$ and also backwards to $k=2 \times 10^{-3} \, {\rm Mpc}^{-1}$, the largest scale on which the primordial power spectrum is well constrained by Planck. No significant changes to our results occur if instead we take $N=0$ to correspond to the smallest $k$ value probed by Planck instead of the pivot scale $k_{0}$.

For each model we calculate the primordial power spectrum of the curvature perturbation as a function of scale using the Stewart-Lyth (SL) formula~\cite{Stewart:1993bc}:
\begin{equation}
\label{sl}
{\cal P}_{\cal R}^{\rm SL}(k) = \left. \frac{ \left[ 1 - (2C+1) \epsilon_H + C (^1\lambda_H) \right]^2}{\pi \epsilon_H} \left( \frac{H}{m_{\rm Pl}} \right)^2   \right \rvert_{k=a H}  \,.
\end{equation}
where $C= -2 + \ln{2} + \gamma \approx -0.729$ with $\gamma$ the Euler-Mascheroni constant. 
This formula is derived via a slow-roll expansion around the exact solution for power-law inflation (where the scale factor varies with time as $a \propto t^{p}$ with $p>1$ and the slow roll parameters are constant). A full numerical evaluation of the Mukhanov-Sasaki equation~\cite{Mukhanov:1985rz,Sasaki:1986hm} finds that scales that exit the horizon very close to the end of inflation~\cite{Leach:2000yw,Josan:2010cj} are amplified relative to the SL formula. However this difference is small compared to the difference between the SL formula and power-law extrapolations of the power spectrum and the SL formula otherwise accurately matches the results of a full numerical calculation for the models generated by the flow formalism~\cite{Josan:2010cj}.

\begin{figure}[t]
\includegraphics[width=0.45\textwidth]{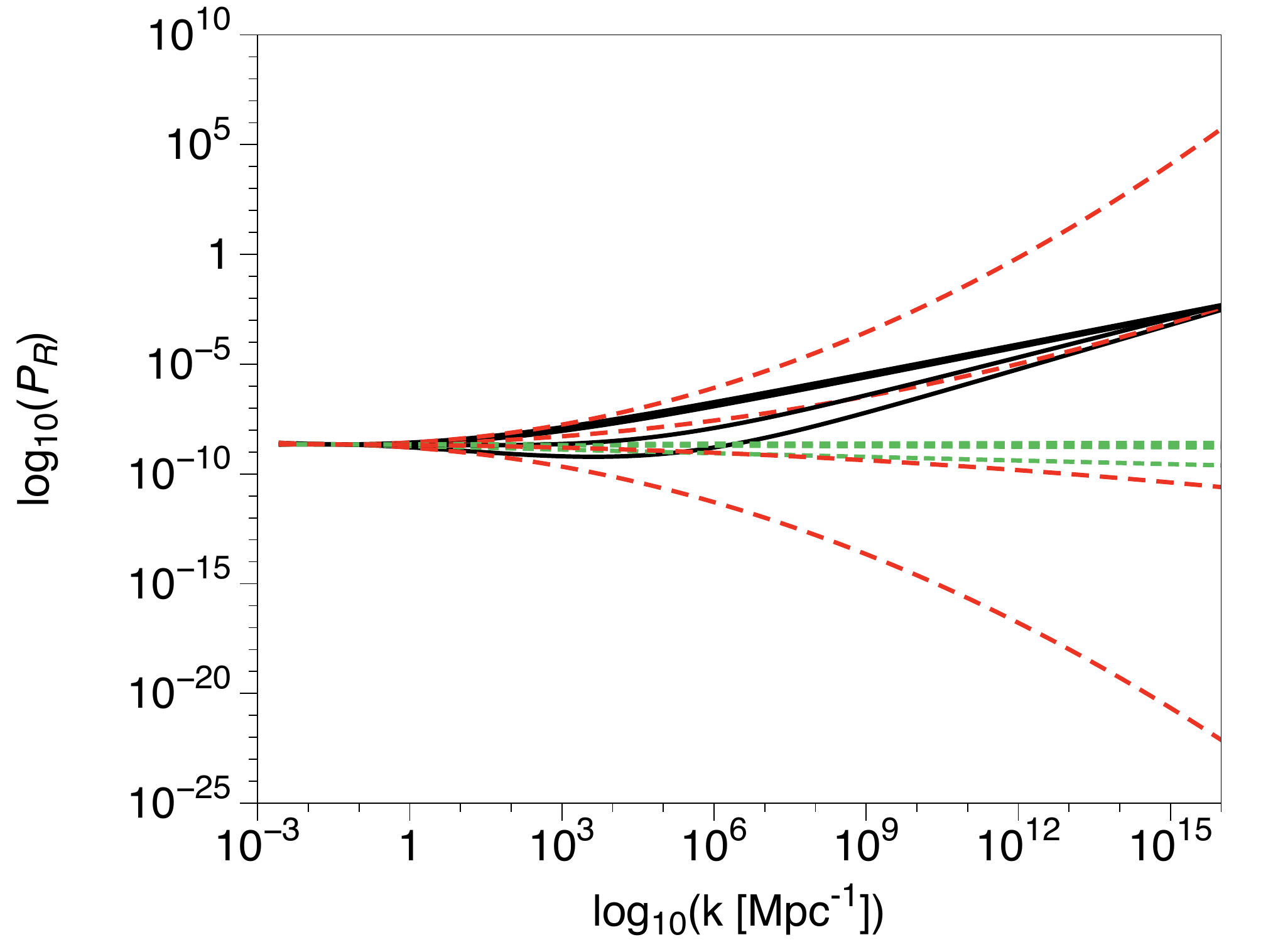}\\
\includegraphics[width=0.45\textwidth]{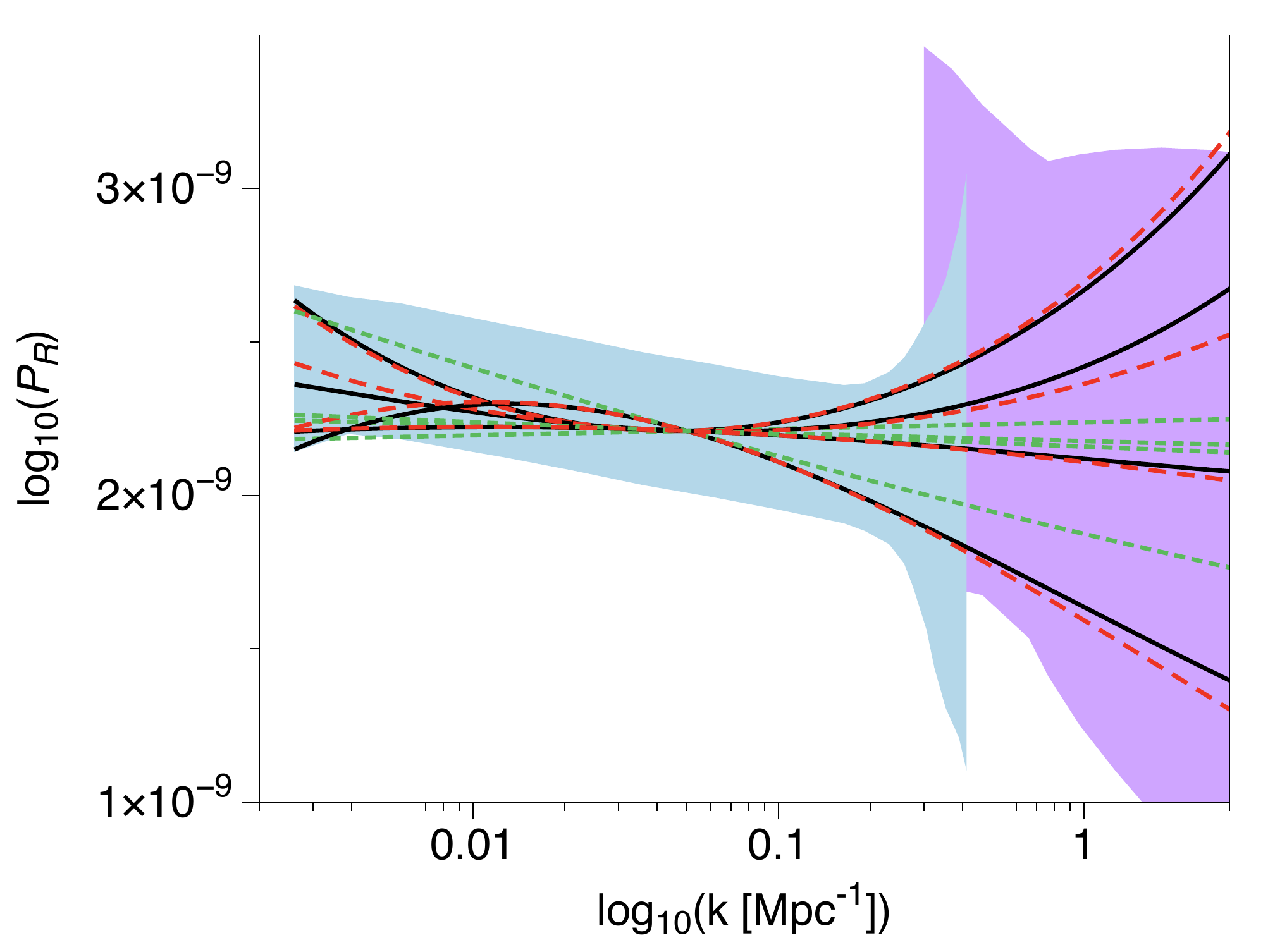}
\caption{\label{fig:flow} Four examples of the primordial power spectrum of the curvature perturbation, ${\cal P}_{\cal R}(k)$, generated using the flow formalism. The `initial' values of the Hubble slow roll parameters have been chosen to give ${\cal P}_{\cal R}(k= 10^{16} \, {\rm Mpc}^{-1}) \approx 10^{-2}$ while satisfying the constraints from Planck and the Lyman-$\alpha$ forest on large scales (see text for further details). The black solid lines are calculated using the Stewart-Lyth expression, Eq.~(\ref{sl}), while the red long-dashed and green short-dashed lines are from Taylor expansions of the spectral index at 2nd and 1st order respectively, with the spectral index and its running calculated using the initial values of the slow roll parameters. The upper panel shows the full range of scales considered while the lower panel is restricted to the scales constrained by the CMB and Large Scale Structure (LSS). The blue and purple shaded regions in the lower panel show the 2$\sigma$ Bayesian reconstruction of the primordial power spectrum from Planck 2015 (Fig.~26 of Ref.~\cite{Ade:2015lrj}) and the Lyman-$\alpha$ forest (Fig.~7 of Ref.~\cite{Bird:2010mp}) respectively.}
\end{figure}

We select models for which 
\begin{itemize}
\item {${\cal P}_{\cal R}^{SL}(k)$ on large scales ($2 \times 10^{-3} \, {\rm Mpc}^{-1} < k < 3 \,  {\rm Mpc}^{-1}$) lies within the $2 \sigma$ Bayesian reconstructions of the primordial power spectrum from the Planck 2015 data (Fig.~26 of Ref.~\cite{Ade:2015lrj}) and the Lyman-$\alpha$ forest (Fig.~7 of Ref.~\cite{Bird:2010mp}),}
\item{ the tensor-to-scalar ratio (e.g. Ref.~\cite{Huang:2006yt})
\begin{equation}
 r= 16 \epsilon_{H} + 32 C (\epsilon_{H})^2 - 32 C \epsilon_{H} (^1\lambda_H) \,,
\end{equation}
calculated using the initial values of the slow roll parameters satisfies $r_{0}<0.1$~\cite{Ade:2015lrj},}
\item{the primordial power spectrum on small scales in large enough to produce an interesting abundance of PBHs: ${\cal P}_{\cal R}(k_{\rm ref}) \approx 10^{-2}$.}
\end{itemize}
In these cases we compare the Stewart-Lyth calculation of the power spectrum with the 1st and 2nd order Taylor expansions of $n_{\rm s}(k)$.  The spectral index and running at the pivot scale are calculated using the initial values of the slow roll parameters and the 2nd order slow roll expressions, e.g. Ref.~\cite{Huang:2006yt},: 
\begin{eqnarray}
n_{{\rm s}} &=& 1 - 4  \epsilon_H + 2  (^1\lambda_H) - 2 C (^2\lambda_H)  - (8C +1)  (\epsilon_H)^2  \nonumber \\
 &+& (6 + 10C)  \epsilon_H    (^1\lambda_H) \,, \\
\frac{{\rm d} n_{\rm s}}{{\rm d}{\rm{ln}}{k}}
 &=&  - 8 (\epsilon_H)^2 + 10 \epsilon_H (^1\lambda_H)  -2 (^2\lambda_H)  \nonumber \\
&  -& (8 + 14C) \epsilon_H   (^2\lambda_H)  + 2C (^1\lambda_H) (^2\lambda_H)  \nonumber \\
&+&  2 C (^3\lambda_H)
 \,.
\end{eqnarray}

The power spectra for four example models are displayed in Fig.~\ref{fig:flow}. Three of these models have been chosen 
to span the range of deviations between the power spectrum calculated using the 2nd order expansion of $n_{\rm s}(k)$ and the SL calculation. As in Sec.~\ref{subsec:pl} we see that the 1st order power spectra decrease weakly with increasing $k$ and are many orders of magnitude smaller on small scales than the SL power spectra, and far too small to form a significant abundance of PBHs.  In this case the 2nd order power spectra can be many orders of magnitude smaller or larger than the SL calculation, or very similar to it. Typically the deviation is non-negligible. 

The case where there is agreement between the 2nd order and SL calculation at $k_{\rm ref}$ is in fact accidental. As discussed earlier, the models where the power spectrum grows with decreasing $k$ have $\epsilon_{\rm H}$ decreasing with increasing $k$ so that the power-spectrum tends to a pure power-law. However the initial value of the running is not negligible so the 2nd order power spectrum isn't a pure power-law.  Therefore if the power spectra were continued to smaller $k$ the 2nd order power spectra would deviate from the SL calculation.
Deviations of many orders of magnitude between the 2nd order power spectra and the SL calculation are typical.
It is not possible to make a concrete, quantitative statement about the size of the deviation between the 2nd order power spectra and the SL calculation.  If we had generated a larger selection of models they would have included models with larger deviations, but this would essentially be due to fine-tuning of the initial slow roll parameters.  Furthermore, as pointed out in Refs.~\cite{Liddle:2003py,Vennin:2014xta}, the flow formalism does not generate an unbiased selection of inflation models. 

The majority of models which have large fluctuations on small scales, while satisfying the CMB and LSS constraints, have power spectra on large scales which are non-monotonic. By fine-tuning the initial slow roll parameters, so that $\epsilon_{H,0}$, $\sigma_{H,0}$ and $(^2\lambda_{H,0}$) are all small, it is possible to produce monotonic behaviour on large scales and still have large fluctuations on small scales. An example of this is also plotted in Fig.~\ref{fig:flow}. In this case the tilt and running at the pivot scale are both small and hence the 2nd order power spectrum behaves in a similar way to the 1st order power spectrum, and is many orders of magnitude smaller than the SL power spectrum on small scales.

We again caution against interpreting these results as demonstrating that PBH formation is natural or generic. In addition to the limitations of the flow formalism mentioned above, fine-tuning of the initial slow roll parameters is required to produce a power-spectrum on small scales which is large enough to produce an interesting number of PBHs while satisfying the CMB and Lyman-$\alpha$ forest constraints on large scales. Furthermore, since the models which produce large fluctuations on small scales have $\epsilon_{\rm H} \rightarrow 0$ at late times, following~Ref.~\cite{Peiris:2008be} we have assumed that an unspecified auxiliary mechanism terminates inflation.

\section{Discussion}
\label{sec:discuss}

We have studied the accuracy of a power-law parameterisation of the primordial power spectrum of the curvature perturbation, Eq.~(\ref{plps}), when studying Primordial Black Hole formation. We first examined how the amplitude on small scales varies when the Taylor expansion of the spectral index, $n_{\rm s}(k)$, is truncated at different orders, using values for the co-efficients of the expansion consistent with the measurements of the tilt, running and running of the running on large scales from the Planck measurements of the anisotropies in the CMB~\cite{Ade:2015lrj}. We found the amplitude typically changes by many orders of magnitude when going from 1st order (only including the tilt) to 2nd order (adding in the running) and also when going from 2nd order to 3rd order (adding in the running of the running). This illustrates that a power-law expansion is not valid when extrapolating over the many orders of magnitude between the scales probed by the CMB and those on which PBHs may form.

Since the CMB probes a relatively small range of scales the errors in the running of the running from Planck are large, and values larger than typically expected from slow roll inflation are allowed. We therefore then used the flow formalism~\cite{Kinney:2002qn} to generate a large ensemble of inflation models, including some where the power spectrum is consistent with the CMB and LSS measurements on large scales and grows sufficiently with increasing $k$ to form PBHs on small scales. In these cases we compared the 1st and 2nd order power-law parameterisations of the power spectrum with the  calculation using the Stewart-Lyth expression~\cite{Stewart:1993bc}. Again we found that the power-law expansion led to amplitudes which were typically incorrect by many orders of magnitude, and could be larger or smaller than the SL value. Since the abundance of PBHs formed depends exponentially on the amplitude of the fluctuations, Eq.~(\ref{beta}), these large errors in the amplitude of the fluctuations lead to huge errors in the abundance of PBHs. We do not quote a value for the size of the error, since it depends significantly on the initial values of the slow roll parameters. For concreteness we focussed on a wavenumber $k_{\rm ref}= 2 \times 10^{16} \, {\rm Mpc}^{-1}$, corresponding to the smallest mass PBHs which do not evaporate by the present day, $M_{\rm PBH} \approx 5 \times 10^{14} \, {\rm g}$~\cite{MacGibbon:1991vc}. However, as can be seen in Fig.~\ref{fig:flow}, the deviations are still large even for smaller $k$, or equivalently heavier PBHs.

It has been pointed out that the flow formalism does not probe the full range of possible inflation models; it effectively assumes a particular form for the potential~\cite{Liddle:2003py} and only samples restricted sections of the potential~\cite{Vennin:2014xta}. Nonetheless these models, where the potential/Hubble parameter does not have a feature at a particular scale and the power spectrum varies gradually, are a best case scenario for a power-law parameterisation of the power spectrum. In models  where PBHs are produced from a spike or peak in the power spectrum at a particular scale (for instance from a local minimum in the potential~\cite{Motohashi:2017kbs,Ballesteros:2017fsr} or multiple fields, see Ref.~\cite{Sasaki:2018dmp})  a power-law extrapolation from cosmological scales is clearly inappropriate.

Our conclusion is that a power-law extrapolation of the primordial power spectrum from cosmological scales down to the small scales on which PBHs may form is invalid. It produces extremely inaccurate values for the amplitude of fluctuations on small scales, and hence the abundance of PBHs formed. As discussed in Sec.~\ref{subsec:flow} these results should not be interpreted as showing that PBHs can be naturally formed in single-field inflation models. Finding flow models where the power spectrum on small scales is large enough to produce an interesting number of PBHs, while satisfying the observational constraints on large scales, requires fine-tuning of the initial slow roll parameters. Furthermore, since the approach we have followed~\cite{Peiris:2008be} assumes that a auxiliary mechanism terminates inflation, these are not strictly speaking single field models.

\vspace*{1cm}
\acknowledgments

A.M.G.  acknowledges  support  from  STFC  grant ST/P000703/1. This research was supported in part by the National Science Foundation under Grant No. NSF PHY11-25915.


\begin{thebibliography}{99}

\bibitem{Bird:2016dcv}
  S.~Bird, I.~Cholis, J.~B.~Muñoz, Y.~Ali-Haïmoud, M.~Kamionkowski, E.~D.~Kovetz, A.~Raccanelli and A.~G.~Riess,
  Phys.\ Rev.\ Lett.\  {\bf 116} (2016) no.20,  201301
  [arXiv:1603.00464 [astro-ph.CO]].

\bibitem{Carr:2016drx}
  B.~Carr, F.~K\"uhnel and M.~Sandstad,
  Phys.\ Rev.\ D {\bf 94} (2016) no.8,  083504
  [arXiv:1607.06077 [astro-ph.CO]].


\bibitem{Allsman:2000kg}
  R.~A.~Allsman {\it et al.} [Macho Collaboration],
  Astrophys.\ J.\  {\bf 550} (2001) L169
  [astro-ph/0011506].

\bibitem{Tisserand:2006zx}
  P.~Tisserand {\it et al.} [EROS-2 Collaboration],
  Astron.\ Astrophys.\  {\bf 469} (2007) 387
  [astro-ph/0607207].

  \bibitem{Mediavilla:2017bok}
  E.~Mediavilla, J.~Jiménez-Vicente, J.~A.~Muñoz, H.~Vives-Arias and J.~Calderón-Infante,
  Astrophys.\ J.\  {\bf 836} (2017) no.2,  L18
  [arXiv:1702.00947 [astro-ph.GA]].


\bibitem{Zumalacarregui:2017qqd}
  M.~Zumalacarregui and U.~Seljak,
  arXiv:1712.02240 [astro-ph.CO].

\bibitem{Monroy-Rodriguez:2014ula}
  M.~A.~Monroy-Rodríguez and C.~Allen,
  Astrophys.\ J.\  {\bf 790} (2014) no.2,  159
  [arXiv:1406.5169 [astro-ph.GA]].


\bibitem{Brandt:2016aco}
  T.~D.~Brandt,
  Astrophys.\ J.\  {\bf 824} (2016) no.2,  L31
  [arXiv:1605.03665 [astro-ph.GA]].

\bibitem{Koushiappas:2017chw}
  S.~M.~Koushiappas and A.~Loeb,
  Phys.\ Rev.\ Lett.\  {\bf 119} (2017) no.4,  041102
  [arXiv:1704.01668 [astro-ph.GA]].

\bibitem{Gaggero:2016dpq}
  D.~Gaggero, G.~Bertone, F.~Calore, R.~M.~T.~Connors, M.~Lovell, S.~Markoff and E.~Storm,
  Phys.\ Rev.\ Lett.\  {\bf 118} (2017) no.24,  241101
  [arXiv:1612.00457 [astro-ph.HE]].
  
  \bibitem{Ali-Haimoud:2016mbv}
  Y.~Ali-Haïmoud and M.~Kamionkowski,
  Phys.\ Rev.\ D {\bf 95} (2017) no.4,  043534
  [arXiv:1612.05644 [astro-ph.CO]].
  
  
  \bibitem{Inoue:2017csr}
  Y.~Inoue and A.~Kusenko,
  JCAP {\bf 1710} (2017) no.10,  034
  [arXiv:1705.00791 [astro-ph.CO]].
 
 \bibitem{Niikura:2017zjd}
  H.~Niikura {\it et al.},
  arXiv:1701.02151 [astro-ph.CO].
 
 \bibitem{Capela:2013yf}
  F.~Capela, M.~Pshirkov and P.~Tinyakov,
  Phys.\ Rev.\ D {\bf 87} (2013) no.12,  123524
  [arXiv:1301.4984 [astro-ph.CO]].

 \bibitem{Inomata:2017vxo}
  K.~Inomata, M.~Kawasaki, K.~Mukaida and T.~T.~Yanagida,
  Phys.\ Rev.\ D {\bf 97} (2018) no.4,  043514
  [arXiv:1711.06129 [astro-ph.CO]].
 
 
 \bibitem{Kawasaki:2016pql}
  M.~Kawasaki, A.~Kusenko, Y.~Tada and T.~T.~Yanagida,
  Phys.\ Rev.\ D {\bf 94} (2016) no.8,  083523
  [arXiv:1606.07631 [astro-ph.CO]].
 
 
 \bibitem{Carr:1974nx}
  B.~J.~Carr and S.~W.~Hawking,
  Mon.\ Not.\ Roy.\ Astron.\ Soc.\  {\bf 168} (1974) 399.
 
 \bibitem{Carr:1975qj}
  B.~J.~Carr,
  Astrophys.\ J.\  {\bf 201} (1975) 1.
 
 
 \bibitem{Carr:1993aq}
  B.~J.~Carr and J.~E.~Lidsey,
  Phys.\ Rev.\ D {\bf 48} (1993) 543.
 
 \bibitem{Sasaki:2018dmp}
  M.~Sasaki, T.~Suyama, T.~Tanaka and S.~Yokoyama,
  Class.\ Quant.\ Grav.\  {\bf 35} (2018) no.6,  063001
  [arXiv:1801.05235 [astro-ph.CO]].
 
   \bibitem{Josan:2009qn}
  A.~S.~Josan, A.~M.~Green and K.~A.~Malik,
  Phys.\ Rev.\ D {\bf 79} (2009) 103520
  [arXiv:0903.3184 [astro-ph.CO]].

 
 \bibitem{Bringmann:2011ut}
  T.~Bringmann, P.~Scott and Y.~Akrami,
  Phys.\ Rev.\ D {\bf 85} (2012) 125027
  [arXiv:1110.2484 [astro-ph.CO]].

\bibitem{Stewart:1996ey}
  E.~D.~Stewart,
  Phys.\ Lett.\ B {\bf 391} (1997) 34
  [hep-ph/9606241].

\bibitem{Stewart:1997wg}
  E.~D.~Stewart,
  Phys.\ Rev.\ D {\bf 56} (1997) 2019
  [hep-ph/9703232].


\bibitem{Leach:2000ea}
  S.~M.~Leach, I.~J.~Grivell and A.~R.~Liddle,
  Phys.\ Rev.\ D {\bf 62} (2000) 043516
  doi:10.1103/PhysRevD.62.043516
  [astro-ph/0004296].


 \bibitem{Peiris:2008be}
  H.~V.~Peiris and R.~Easther,
  JCAP {\bf 0807} (2008) 024
  [arXiv:0805.2154 [astro-ph]].
 
 \bibitem{Josan:2010cj}
  A.~S.~Josan and A.~M.~Green,
  Phys.\ Rev.\ D {\bf 82} (2010) 047303
  [arXiv:1004.5347 [hep-ph]].


 
 \bibitem{Motohashi:2017kbs}
  H.~Motohashi and W.~Hu,
  Phys.\ Rev.\ D {\bf 96} (2017) no.6,  063503
  [arXiv:1706.06784 [astro-ph.CO]].
 
 \bibitem{Ballesteros:2017fsr}
  G.~Ballesteros and M.~Taoso,
  Phys.\ Rev.\ D {\bf 97} (2018) no.2,  023501
  [arXiv:1709.05565 [hep-ph]].


 
 \bibitem{Kosowsky:1995aa}
  A.~Kosowsky and M.~S.~Turner,
  Phys.\ Rev.\ D {\bf 52} (1995) R1739
  [astro-ph/9504071].
 
 
 \bibitem{Ade:2015lrj}
  P.~A.~R.~Ade {\it et al.} [Planck Collaboration],
  Astron.\ Astrophys.\  {\bf 594} (2016) A20
  [arXiv:1502.02114 [astro-ph.CO]].
 
 
 \bibitem{Ade:2013zuv}
  P.~A.~R.~Ade {\it et al.} [Planck Collaboration],
  Astron.\ Astrophys.\  {\bf 571} (2014) A16
  [arXiv:1303.5076 [astro-ph.CO]].



\bibitem{Leach:2002ar}
  S.~M.~Leach, A.~R.~Liddle, J.~Martin and D.~J.~Schwarz,
  Phys.\ Rev.\ D {\bf 66} (2002) 023515
  [astro-ph/0202094].

 
\bibitem{Mollerach:1993sy}
  S.~Mollerach, S.~Matarrese and F.~Lucchin,
  Phys.\ Rev.\ D {\bf 50} (1994) 4835
  doi:10.1103/PhysRevD.50.4835
  [astro-ph/9309054].


\bibitem{Gilbert:1995wg}
  J.~Gilbert,
  Phys.\ Rev.\ D {\bf 52} (1995) 5486.
  doi:10.1103/PhysRevD.52.5486

\bibitem{Vallinotto:2003vf}
  A.~Vallinotto, E.~J.~Copeland, E.~W.~Kolb, A.~R.~Liddle and D.~A.~Steer,
  Phys.\ Rev.\ D {\bf 69} (2004) 103519
  doi:10.1103/PhysRevD.69.103519
  [astro-ph/0311005].

 
\bibitem{Lidsey:1995np}
  J.~E.~Lidsey, A.~R.~Liddle, E.~W.~Kolb, E.~J.~Copeland, T.~Barreiro and M.~Abney,
  Rev.\ Mod.\ Phys.\  {\bf 69} (1997) 373
  [astro-ph/9508078].
 
 
\bibitem{Alabidi:2009bk}
  L.~Alabidi and K.~Kohri,
  Phys.\ Rev.\ D {\bf 80} (2009) 063511
  [arXiv:0906.1398 [astro-ph.CO]].
 
 \bibitem{Carr:2017edp}
  B.~Carr, T.~Tenkanen and V.~Vaskonen,
  Phys.\ Rev.\ D {\bf 96} (2017) no.6,  063507
  [arXiv:1706.03746 [astro-ph.CO]].
 
 \bibitem{Kohri:2018qtx}
  K.~Kohri and T.~Terada,
  arXiv:1802.06785 [astro-ph.CO].
 

 \bibitem{MacGibbon:1991vc}
  J.~H.~MacGibbon and B.~J.~Carr,
  Astrophys.\ J.\  {\bf 371} (1991) 447.

 
 \bibitem{Niemeyer:1997mt}
  J.~C.~Niemeyer and K.~Jedamzik,
  Phys.\ Rev.\ Lett.\  {\bf 80} (1998) 5481
  [astro-ph/9709072].
 
 \bibitem{Franciolini:2018vbk}
  G.~Franciolini, A.~Kehagias, S.~Matarrese and A.~Riotto,
  JCAP {\bf 1803} (2018) no.03,  016
  [arXiv:1801.09415 [astro-ph.CO]].
 
 
 \bibitem{Ando:2018qdb}
  K.~Ando, K.~Inomata and M.~Kawasaki,
  arXiv:1802.06393 [astro-ph.CO].
 
 
 
 \bibitem{Harada:2013epa}
  T.~Harada, C.~M.~Yoo and K.~Kohri,
  Phys.\ Rev.\ D {\bf 88} (2013) no.8,  084051
   Erratum: [Phys.\ Rev.\ D {\bf 89} (2014) no.2,  029903]
  [arXiv:1309.4201 [astro-ph.CO]].
 
  
 
 \bibitem{Hoffman:2000ue}
  M.~B.~Hoffman and M.~S.~Turner,
  Phys.\ Rev.\ D {\bf 64} (2001) 023506
  [astro-ph/0006321].
 
 \bibitem{Kinney:2002qn}
  W.~H.~Kinney,
  Phys.\ Rev.\ D {\bf 66} (2002) 083508
  [astro-ph/0206032].
 
 \bibitem{Salopek:1990jq}
  D.~S.~Salopek and J.~R.~Bond,
  Phys.\ Rev.\ D {\bf 42} (1990) 3936.
 
 
 
 \bibitem{Liddle:2003py}
  A.~R.~Liddle,
  Phys.\ Rev.\ D {\bf 68} (2003) 103504
  [astro-ph/0307286].
 
 \bibitem{Ramirez:2005cy}
  E.~Ramirez and A.~R.~Liddle,
  Phys.\ Rev.\ D {\bf 71} (2005) 123510
  [astro-ph/0502361].
 
 
 
 
 
 
 \bibitem{Stewart:1993bc}
  E.~D.~Stewart and D.~H.~Lyth,
  Phys.\ Lett.\ B {\bf 302} (1993) 171
  [gr-qc/9302019].
 

\bibitem{Mukhanov:1985rz}
  V.~F.~Mukhanov,
  JETP Lett.\  {\bf 41} (1985) 493
   [Pisma Zh.\ Eksp.\ Teor.\ Fiz.\  {\bf 41} (1985) 402].

\bibitem{Sasaki:1986hm}
  M.~Sasaki,
  Prog.\ Theor.\ Phys.\  {\bf 76} (1986) 1036.


 \bibitem{Leach:2000yw}
  S.~M.~Leach and A.~R.~Liddle,
  Phys.\ Rev.\ D {\bf 63} (2001) 043508
  [astro-ph/0010082].
 

\bibitem{Bird:2010mp}
  S.~Bird, H.~V.~Peiris, M.~Viel and L.~Verde,
  Mon.\ Not.\ Roy.\ Astron.\ Soc.\  {\bf 413} (2011) 1717
  [arXiv:1010.1519 [astro-ph.CO]].

 
  \bibitem{Huang:2006yt}
  Q.~G.~Huang,
  Phys.\ Rev.\ D {\bf 76} (2007) 043505
  [astro-ph/0610924].


 \bibitem{Vennin:2014xta}
  V.~Vennin,
  Phys.\ Rev.\ D {\bf 89} (2014) no.8,  083526
  [arXiv:1401.2926 [astro-ph.CO]].

 
\end{thebibliography}
\end{document}